**Ultrasmall $Au_{10–12}(SG)_{10–12}$ Nanomolecules for High Tumor Specificity and Cancer Radiotherapy**

*Xiao-Dong Zhang,\*[†]  Zhentao Luo,[†] Jie Chen, Xiu Shen, Shasha Song, Yuanming Sun, Saijun Fan, Feiyue Fan, David Tai Leong and Jianping Xie\**

[\*]     Dr. X. D. Zhang, [†] J. Chen, Prof. X. Shen, S. S. Song, Prof. Y. M. Sun, Dr. Prof. S. J. Fan and Prof. F. Y. Fan
Tianjin Key Laboratory of Molecular Nuclear Medicine, Institute of Radiation Medicine, Chinese Academy of Medical Sciences and Peking Union Medical College, Tianjin, 300192, China
Email: xiaodongzhang@tju.edu.cn

[\*]     Z. Luo, [†] Dr. Prof. D. T. Leong, and Dr. Prof. J. Xie
Department of Chemical and Biomolecular Engineering,
National University of Singapore, 4 Engineering Drive 4, Singapore 117585 (Singapore)
Email: chexiej@nus.edu.sg

[ † ] These authors contributed equally to this work.

Keywords: Gold nanoclusters; Tumor uptake; Radiotherapy; Gold nanomolecules; Radiosensitive

**Note: The tumor uptake of $Au_{10–12}(SG)_{10–12}$ is about 50 % ID/g after 12-72 h.**

Radiotherapy has been considered as part of the treatment regime following tumor surgical removal. It has high efficacy as up to 50% patients are treated with radiotherapy during their battle against cancer.[1] Nevertheless, high energy radiation during the treatment not only kills tumor cells but also obliterates healthy cells along with them, leading to unavoidable damage to normal tissues. Localizing and controlling the radiation dose can maximize tumor eradication and minimize side effects. The use of radiosensitizers to increase the local treatment efficacy under a relatively low and safe radiation dose is the most promising solution to address this challenge. Conventional drug based radiosensitizers like misonidazole while are efficacious at the site of the tumor do not have any targeting capabilities and rely heavily on precise localization of the drug to the tumor cells.[2] Sometimes for very small tumors with dispersed distribution within a tissue, it becomes impossible to avoid the interspersed normal tissues while only affecting the tumor cells. An ideal radiosensitizer should have high radiotherapy enhancement, good tumor targeting capability (at both tissue and cellular level), good biocompatibility (or low toxicity), and efficient renal clearance to avoid potential short- and long-term detrimental effects on the patient. No radiosensitizers in the current development can meet all these requirements.

To address these unresolved challenges, here we report a new class of radiosensitizers – several gold (Au) atoms embedded inside a peptide shell comprising of a naturally-occurring peptide such as glutathione or GSH. As illustrated in Scheme 1, the as-designed nanomolecule has a well-defined molecular formula of $Au_{10-12}(SG)_{10-12}$, and can be classified as GSH-Au nanomolecule or nanocluster.[3] The GSH-covered surface of the nanomolecule has similar physicochemical and physiological properties as that of a polypeptide comprising of several GSH molecules, and this class of materials have been shown to have good biocompatibility in biological systems.[4] The GSH shell also makes good tumor deposition of the GSH-Au nanomolecules possible by affecting their in vivo pharmacokinetics. The ultrasmall size and the peptide shell may help the GSH-Au nanomolecules escape the reticulo-endothelial system (RES), thus improving their deposition in tumor. Moreover, the high ratio of GSH in the as-designed GSH-Au nanomolecules (or highly exposed GSH on the nanomolecule surface) could possibly activate the GSH transporters inside the body, which could facilitate the uptake of GSH-Au nanomolecules by tumor cells, thereby improving their accumulation inside the tumor.

The Au atoms in the GSH-Au nanomolecules could be used as radiosensitizers to enhance the therapeutic efficiency of radiotherapy. Among several high-Z materials recently developed as radiosensitizers,[5] including iodine (Z = 53), and gadolinium (Z = 64), gold is unmatched in terms of enhancement efficiency because of its large atomic number of 79. The several Au atoms embedded inside a GSH shell could further improve the therapeutic efficiency by providing a locally high Au concentration in tumor with potential synergistic effects. The radiotherapy enhancement comes from the direct interaction between Au and radiation, where, upon receiving high-energy X-ray or gamma-ray radiation, the GSH-Au nanomolecules become a new source of radiation and emit high energy through scattered photons, photoelectrons, Compton electrons, Auger electrons, and electron-positron pairs, causing radiochemicals (free radicals and ionizations) within the cells that can damage and kill cancer cells.[6]

Taken together, the GSH-Au nanomolecules inherit attractive features of both gold atoms and naturally occurring molecules, which could prolong their blood circulation and improve their tumor deposition. The ultrasmall size of GSH-Au nanomolecules may also make high renal clearance possible after treatment, which could minimize their potential side effects due to the accumulation of Au in body. This feature could not be replicated by most of other inorganic-based theranostic agents, such as metal nanoparticles,[7] carbon nanotubes,[8] and semiconductor quantum dots,[9] because they have relatively large hydrodynamic diameters (HDs, typically >10 nm), which are above the threshold of kidney filtration (~5.5 nm).[10] The metabolizable feature of the GSH-Au nanomolecules may further pave their way towards the clinical uses.

The preparation of GSH-Au nanomolecules is simple (Supporting Information).[11] In a typical synthesis, aqueous solutions of GSH and HAuCl$_4$ (GSH-to-Au ratio = 2:1) were first mixed at 25 °C for 5 min, followed by the addition of NaOH to bring the pH of the mixture to ~7.0. The mixture was then incubated at 40 °C for ~2 h, leading to the formation of the GSH-Au nanomolecules. The above formation process involved two steps. The first step was the reduction of Au(III) by GSH to form insoluble aggregates of polymeric Au(I)−SG complexes, and the second step was initiated by the addition of NaOH, which caused the dissociation of the polymeric complexes to form soluble oligomeric Au(I)−SG complexes or GSH-Au nanomolecules. The resulting reaction mixture was clear and colorless. The as-synthesized GSH-Au nanomolecules showed two distinct peaks at 330 and 375 nm (Figure 1a), which matched nicely with $Au_{10-12}(SG)_{10-12}$ nanomolecules reported by Negishi et al.[12] The formation of $Au_{10-12}(SG)_{10-12}$ nanomolecules in the product was confirmed by its electrospray ionization mass spectrum (Figure 1b–d).

The as-prepared $Au_{10-12}(SG)_{10-12}$ nanomolecules showed ultrahigh uptake in tumor compared to normal tissues; such targeting capability is essential for minimizing damage to normal tissues during the radiation treatment. The tumor uptake of $Au_{10-12}(SG)_{10-12}$ nanomolecules was first investigated by analyzing the standardized uptake values (SUVs) of Au in the tumor tissue at different time points post injection (p.i.). SUV is defined as (weight of Au/weight of tissue sample)/(weight of Au injected into animal/total body weight). The mice were intraperitoneally injected with $Au_{10-12}(SG)_{10-12}$ nanomolecules, and the postmortem tissue samples were treated and analyzed (Supplementary Information). As shown in Figure 2a, the accumulation of Au in the tumor increased sharply from 6 to 10 h p.i., and gradually reached a maximum at 24 h p.i with a SUV of 10.86, which could be retained up to 48 h p.i. The tumor uptake of $Au_{10-12}(SG)_{10-12}$ nanomolecules at 24 h p.i. is about an order of magnitude higher than those of tiopronin-protected Au nanoparticles (~2 nm) and PEG-coated Au nanorods (~20 nm).[13] The ultrahigh uptake of $Au_{10-12}(SG)_{10-12}$ nanomolecules in tumor could be attributed to two of their structural and size features.[14] The first feature is their ultrasmall HD, which is around 2 nm. Particles in this size regime may benefit most from the enhanced permeability and retention (EPR)[15] and nanomaterials-induced endothelial cell leakiness (NanoEL) effects.[16] The second feature is their high GSH-to-Au ratio, which is 1:1. A high content of GSH in the as-designed molecules could largely lead them to behave like a polypeptide consisting of GSH molecules, which could help the nanomolecules escape the RES absorption, as well as may activate the GSH transporters on cell surface to further improve the uptake of $Au_{10-12}(SG)_{10-12}$ nanomolecules in tumor cells.

To further understand the reason for the ultrahigh tumor uptake of $Au_{10-12}(SG)_{10-12}$ nanomolecules, we studied the pharmacokinetics of intraperitoneally injected $Au_{10-12}(SG)_{10-12}$ nanomolecules in mice. As shown in Figure 2b, it is obvious that $Au_{10-12}(SG)_{10-12}$ nanomolecules followed a two-compartment pharmacokinetics. It has a distribution half-life (first phase $t_{1/2\alpha}$) of ~2.4 h, which is slightly longer than the previously reported $Au_{25}(SG)_{18}$ nanomolecules and other small Au particles.[4b, 17] On the other hand, Figure 2b also suggests that $Au_{10-12}(SG)_{10-12}$ nanomolecules have a blood-elimination half-life (second phase $t_{1/2\beta}$) of ~22 h. Moreover, it was found that, even after 24 h p.i., the blood concentration of $Au_{10-12}(SG)_{10-12}$ nanomolecules was still above 4.91 SUV. This value is ~20 times higher than that of the $Au_{25}(SG)_{18}$ nanomolecules, and higher than reported small Au NPs.[4b, 17] Taken together, the combination of both the long half-lives and the long-lasting high blood concentration could be the major contributor for the ultrahigh tumor uptake of $Au_{10-12}(SG)_{10-12}$ nanomolecules.

To further understand the in vivo behavior of $Au_{10-12}(SG)_{10-12}$ nanomolecules, we have studied their biodistribution at 24 h and 23 days p.i. As shown in Figure 3, at 24 h p.i., the concentration of $Au_{10-12}(SG)_{10-12}$ nanomolecules in tumor was much higher than that of all other key organs including kidney and liver. For instance, the ratios of the nanomolecule concentration in tumor to that in kidney and liver were 1:0.172 and 1:0.0446, respectively. Such a high targeting specificity of $Au_{10-12}(SG)_{10-12}$ nanomolecules is highly desirable because it can constrain the radiotherapy sensitization within the tumors and minimize possible damages to normal tissues. The high targeting specificity of $Au_{10-12}(SG)_{10-12}$ nanomolecules could be attributed to their biocompatible surface and ultrasmall size features, which help them evade the uptake by the RES organs, such as liver and spleen. On the other hand, at 23 days p.i., the nanomolecule concentrations in all the key organs and the tumor were dropped below 0.019 SUV, which clearly suggest that $Au_{10-12}(SG)_{10-12}$ nanomolecules are highly renal clearable.

To confirm the selective deposition of $Au_{10-12}(SG)_{10-12}$ nanomolecules in tumor, X-ray computed tomography (CT) was used to image the distribution of the nanomolecules in body. In vivo X-ray CT imaging is a non-invasive and reliable method for tumor imaging.[4a, 18] The CT signal depends on the concentration of Au atoms (from the injected $Au_{10-12}(SG)_{10-12}$ nanomolecules) in tissues, and a CT value of 745 HU that corresponds to 40 mM of Au (Figure S1) is suitable for in vivo imaging. $Au_{10-12}(SG)_{10-12}$ nanomolecules (40 mM Au, 0.2 mL) were intravenously injected into mice, and three- and two-dimensional X-ray CT images were recorded. As shown in Figure 4, an obvious tumor uptake was clearly seen at the tumor site (indicated by arrows) at 6 h p.i. The corresponding CT value was determined to be 326 HU, which is significantly higher than that of the muscle tissue (207 HU). In addition, a clear boundary between the tumor and normal tissues was observed. Taken together, the X-ray CT images provide strong evidences on the above discussed biodistribution data of $Au_{10-12}(SG)_{10-12}$ nanomolecules.

The radiotherapy sensitization efficacy of the GSH-Au nanomolecules were tested using U14 tumor bearing nude mice as the animal model. The mice were intraperitoneally injected with $Au_{10-12}(SG)_{10-12}$ nanomolecules (10 mM Au, 0.2 mL) to a concentration of 20 mg-Au/kg-body. The mice were irradiated under $^{137}$Cs gamma radiation of 3600 Ci at a 5 Gy dose at 24 h p.i. when the tumor uptake of $Au_{10-12}(SG)_{10-12}$ nanomolecules reached the maximum (Figure 2a). The time-dependent tumor volumes and tumor weights at the time point of 23 day in the sacrificed mice were measured (Figure 5a and 5b). The tumor volume did not show any decrease for mice treated with $Au_{10-12}(SG)_{10-12}$ only. As compared with the control group and mice treated with radiation only, tumor volume in mice treated with both $Au_{10-12}(SG)_{10-12}$ nanomolecules and radiation decreased ~65% and ~57%, respectively. Correspondingly, the tumor weight in mice treated with both $Au_{10-12}(SG)_{10-12}$ nanomolecules and radiation decreased significantly relative to tumors in the control group, mice treated with $Au_{10-12}(SG)_{10-12}$ only, and mice treated with radiation only. Therefore, the as-designed $Au_{10-12}(SG)_{10-12}$ nanomolecules can significantly enhance the radiotherapy efficiency.

The extent of the radiotherapy sensitization effect strongly depends on the tumor uptake of the radiosensitizers. The traditional molecular radiosensitizers such as cisplatin can achieve high tumor uptake as well as good radiotherapy sensitization. However, the renal clearance of cisplatin was slow, which could cause potential kidney toxicity.[19] On the other hand, those radiosensitizers with relatively large HDs (e.g., >20 nm) were not able to evade RES clearance, and they may also cause potential liver toxicity.[20] However, the as-designed $Au_{10-12}(SG)_{10-12}$ nanomolecules in this study feature with efficient tumor uptake, high targeting specificity, and efficient renal clearance. As an attractive potential radiosensitizer, the toxicity response of $Au_{10-12}(SG)_{10-12}$ nanomolecules, including blood chemistry, biochemistry and pathology, were further examined. Loss of the body weight (Figure S2a) or abnormal organ indices (Figure S2b–c) were observed for mice treated with $Au_{10-12}(SG)_{10-12}$ nanomolecules at a dose of 20 mg-Au/kg-body, which was used for radiotherapy sensitization. The typical chemistry and biochemistry data (Figure S3 and S4) showed that platelets (PLT), hematocrit (HCT), and blood urea nitrogen (BUN) were decreased in mice treated with $Au_{10-12}(SG)_{10-12}$ nanomolecules, but they were recovered at 23 days p.i. No obvious damage to key organs including the liver, spleen, and kidney were observed in mice treated with $Au_{10-12}(SG)_{10-12}$ nanomolecules (Figure S5). Therefore, $Au_{10-12}(SG)_{10-12}$ nanomolecules can be considered safe for the radiotherapy with the current doses.

In summary, the ultrasmall size of the $Au_{10-12}(SG)_{10-12}$ nanomolecules could increase the tumor uptake and targeting specificity via the improved EPR effect, while the highly exposed biocompatible GSH shell on the nanomolecule surface could further contribute to their tumor uptake by allowing the nanomolecules to escape the RES absorption and activating the transporter on cell surface. The ultrahigh tumor uptake, targeting specificity, and efficient renal clearance of ultrasmall $Au_{10-12}(SG)_{10-12}$ nanomolecules with highly exposed GSH ligands allows them to be ideal radiotherapy sensitizers that can enhance the safety and efficacy of radiotherapy.

**Experimental Section**

A detailed description of experimental procedures can be found in the Supporting Information.

**Acknowledgment**

This work was financially supported by the National Natural Science Foundation of China (Grant No.81000668), Natural Science Foundation of Tianjin (Grant No. 13JCQNJC13500), the Subject Development Foundation of IRM, CAMS (Grant No.SF1207 and SZ1336), the Foundation of Union New Star, CAMS (No.1256), and PUMC Youth Fund and the Fundamental Research Funds for the Central Universities (Grant No. 3332013043). Work at National University of Singapore was supported by the Ministry of Education, Singapore, under grant R-279-000-409-112.

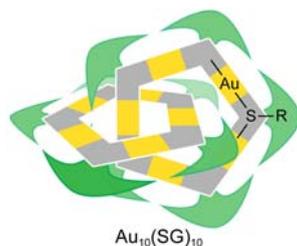

**Scheme 1**. Schematic illustration of the structure of $Au_{10}(SG)_{10}$ nanomolecule.

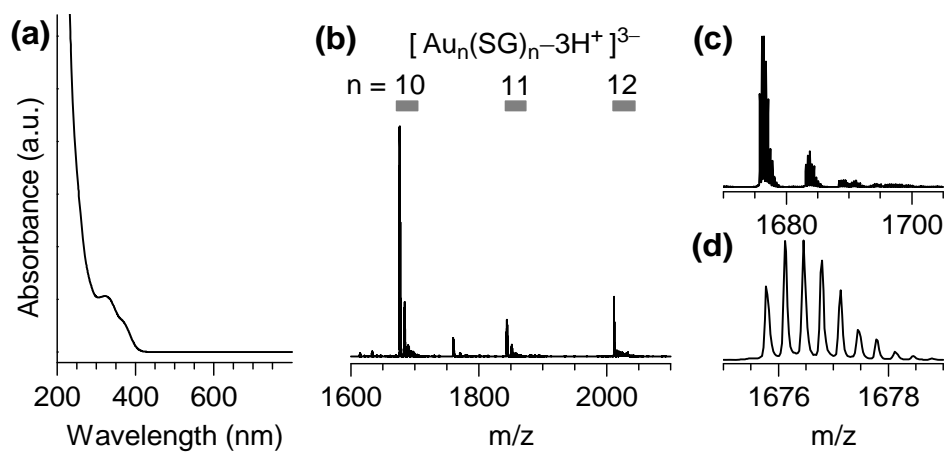

**Figure 1.** (a) UV-vis absorption and (b-d) ESI mass spectrum of the as-synthesized GSH-Au nanomolecules, indicating the formation of $Au_{10-12}(SG)_{10-12}$ nanomolecules in the product. The series of isotope distributions shown in (c) are resulted from the replacement of the carboxyl $H^+$ of GSH by $Na^+$ or $K^+$. The red line in (d) is the simulated isotope distribution of $[Au_{10}(SG)_{10}-3H^+]^{3-}$.

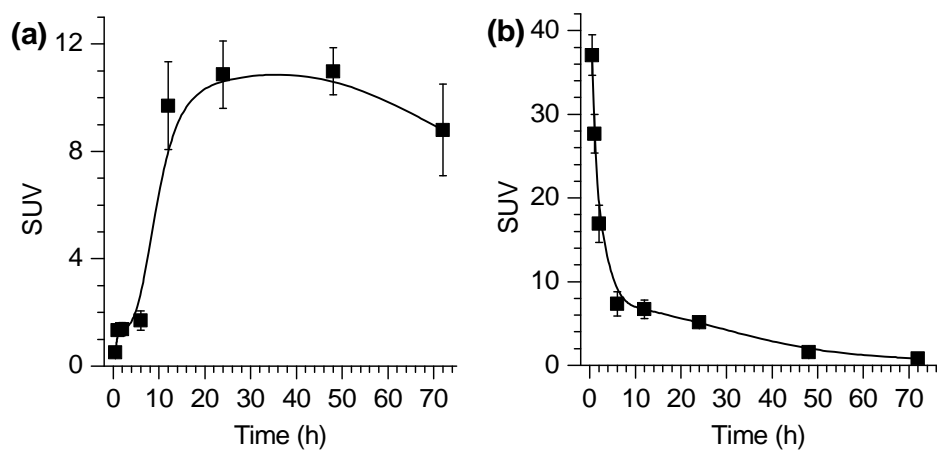

**Figure 2.** (a) Standard uptake values (SUV) of $Au_{10-12}(SG)_{10-12}$ nanomolecules in tumor at different time points p.i. (b) Pharmacokinetics of $Au_{10-12}(SG)_{10-12}$ nanomolecules in mice from 0 to 72 h p.i.

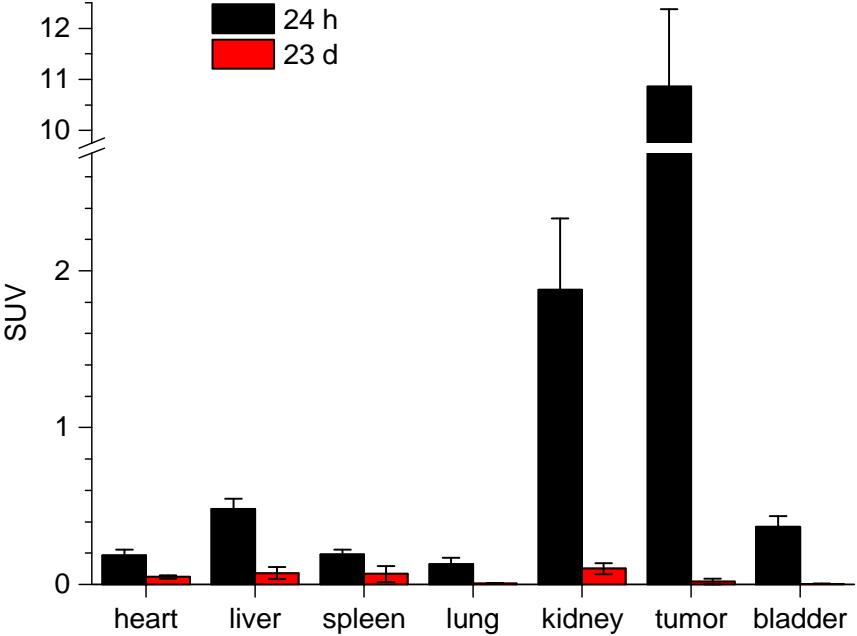

**Figure 3.** Biodistribution of $Au_{10-12}(SG)_{10-12}$ at 24 h and 23 days p.i.

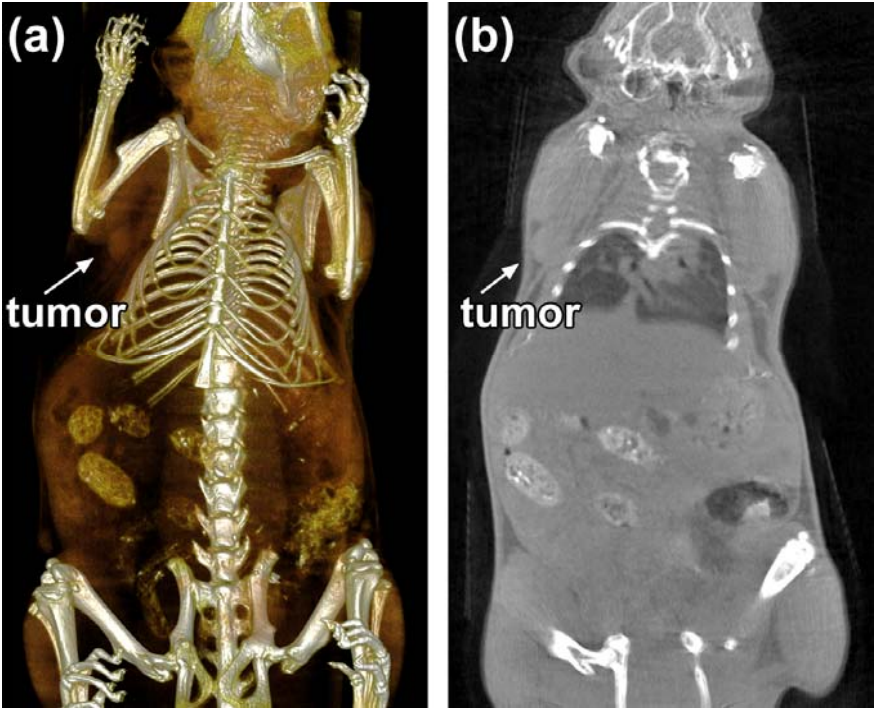

**Figure 4.** (a) Three- and (b) two-dimensional small animal X-ray CT imaging of Au$_{10-12}$(SG)$_{10-12}$ at 6 h p.i.

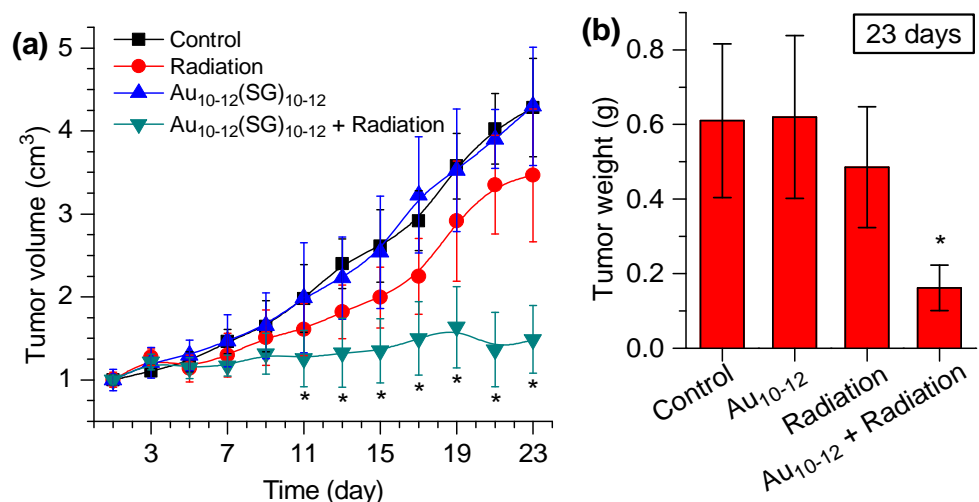

**Figure 5.** Time-course studies of tumor (a) volumes and (b) weights of untreated mice (control), mice treated with Au$_{10-12}$(SG)$_{10-12}$ only, mice treated with radiation only, and mice treated with both Au$_{10-12}$(SG)$_{10-12}$ and radiation. Data is analyzed by student's *t*-test, and the star denotes significant difference from the control group ($p < 0.05$).

# Table of Content

Radiosensitizers can increase the local treatment efficacy under a relatively low and safe radiation dose, thereby facilitating tumor eradication and minimizing side effects. Here, we report a new class of radiosensitizers that contain several gold (Au) atoms embedded inside a peptide shell (e.g., $Au_{10-12}(SG)_{10-12}$) and can achieve ultrahigh tumor uptake (10.86 SUV at 24 h post injection) and targeting specificity, efficient renal clearance, and high radiotherapy enhancement.

**Ultrasmall $Au_{10-12}(SG)_{10-12}$ Nanomolecules for High Tumor Specificity and Cancer Radiotherapy**

*Xiao-Dong Zhang,\* Zhentao Luo, Jie Chen, Xiu Shen, Shasha Song, Yuanming Sun, Saijun Fan, Feiyue Fan, David Tai Leong and Jianping Xie\**

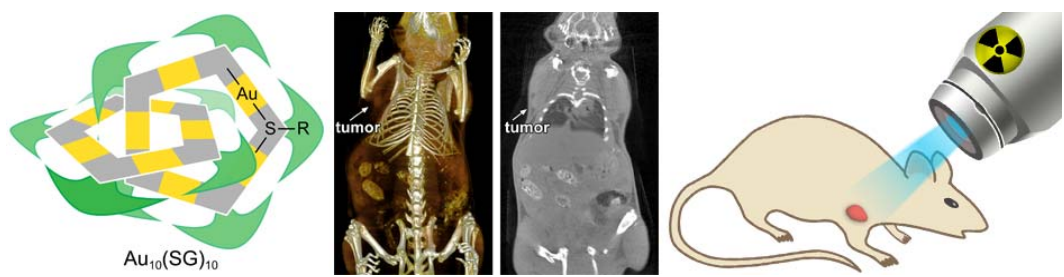

**Ultrasmall Au$_{10-12}$(SG)$_{10-12}$ Nanomolecules for High Tumor Specificity and Cancer Radiotherapy**

*Xiao-Dong Zhang,\* [†]Zhentao Luo, [†] Jie Chen, Xiu Shen, Shasha Song, Yuanming Sun, Saijun Fan, Feiyue Fan, David Tai Leong and Jianping Xie\**

[\*]    Dr. X. D. Zhang, [†] J. Chen, Prof. X. Shen, S. S. Song, Prof. Y. M. Sun, Dr. Prof. S. J. Fan and Prof. F. Y. Fan
Tianjin Key Laboratory of Molecular Nuclear Medicine, Institute of Radiation Medicine, Chinese Academy of Medical Sciences and Peking Union Medical College, Tianjin, 300192, China
Email: xiaodongzhang@tju.edu.cn

[\*]    Z. Luo, [†] Dr. Prof. D. T. Leong, and Dr. Prof. J. Xie
Department of Chemical and Biomolecular Engineering,
National University of Singapore, 4 Engineering Drive 4, Singapore 117585 (Singapore)
Email: chexiej@nus.edu.sg

[ † ] These authors contributed equally to this work.

## 1. Experimental Section

**Synthesis of Au$_{10-12}$(SG)$_{10-12}$ Nanomolecules.** The synthesis of Au$_{10-12}$(SG)$_{10-12}$ nanomolecules followed a reported method.[11] In a typical synthesis, aqueous solutions of HAuCl$_4$ (20 mM, 0.5 mL) and GSH (100 mM, 0.2 mL) were mixed with 4.3 mL of ultrapure water under rigorous stirring (500 rpm) at 25 °C for 5 min. A precipitate was formed and was then dissolved by adjusting pH to ~7.0 with NaOH (0.5 M). The solution was incubated at 40 °C. After 2 h, the product, Au$_{10-12}$(SG)$_{10-12}$ nanomolecule, was collected.

**Characterizations of Au$_{10-12}$(SG)$_{10-12}$ Nanomolecules.** The UV-vis absorption spectrum was recorded in aqueous solutions at 25 °C on a spectrophotometer (UV-1800, Shimadzu). The electrospray ionization mass spectrometry (ESI-MS) was carried out with a Bruker MicroTOF-Q ESI time-of-flight system operating in the negative ion mode (sample injection rate 120 μL·min$^{-1}$; capillary voltage 3 kV; nebulizer 1.5 bar; dry gas 4 L·min$^{-1}$ at 160 °C). GSH-Au nanomolecules was purified for ESI-MS measurement according to the following procedure: 200 μL of as-synthesized Au$_{10-12}$(SG)$_{10-12}$ nanomolecules was first precipitated by mixing with acetic acid (20 μL) and ethanol (1.3 mL); the pellet was then washed with Dimethylformamide and redissolved in water (400 μL).

**In vivo Biodistribution.** Forty-eight mice were purchased, maintained, and handled using protocols approved by the Institute of Radiation Medicine, Chinese Academy of Medical Sciences (CAMS). The U14 tumor models were generated by subcutaneously injecting 2 × 10$^6$ cells (in 50 μL of PBS) into the right shoulders of male nude mice. Au$_{10-12}$(SG)$_{10-12}$ nanomolecules (3 mM per Au atoms, 0.2 mL) was injected through the intraperitoneal routes into mice. The mice were sacrificed at 0.5, 1, 2, 6, 12, 24, 48, and 72 h post injection (p.i.). The tumor and main organs including the liver, kidney, spleen, heart, lung, and brain were collected and digested using a microwave system CEM Mars 5 (CEM, Kamp

Lintfort, Germany). Their Au contents were then measured using inductively coupled plasma mass spectrometry (ICP-MS, Agilent 7500 CE, Agilent Technologies, Waldbronn, Germany).

**In vivo Imaging.** Eighteen mice were purchased, maintained, and handled using protocols approved by the Institute of Radiation Medicine, CAMS. The U14 tumor models were generated by subcutaneously injecting $2 \times 10^6$ cells (in 50 μL of PBS) into the right shoulders of male nude mice. The mice were anesthetized by chloral hydrate before the experiment. For CT imaging, $Au_{10-12}(SG)_{10-12}$ nanomolecules (40 mM per Au atoms, 0.2 mL) was injected through the intraperitoneal routes into the mice. Each mouse was imaged on a small-animal scanner (microPET/CT, Inveon, Siemens) and was exposed to a 10-min CT scan. The images were reconstructed using the filtered back-projection algorithm with CT-based photon-attenuation correction. The CT data were analyzed for regions of interest including the tumor, bladder, and spleen.

**In vivo Radiation Therapy.** Forty-eight mice were purchased, maintained, and handled using protocols approved by the Institute of Radiation Medicine, CAMS. The U14 tumor models were generated by subcutaneously injecting $2 \times 10^6$ cells (in 50 μL of PBS) into the right shoulder of BALB/c mice. The mice were intraperitoneally injected with $Au_{10-12}(SG)_{10-12}$ nanomolecules when the tumor volume reached 100–120 mm$^3$ (7 days after tumor inoculation). For each treatment, $Au_{10-12}(SG)_{10-12}$ nanomolecules (10 mM per Au atoms, ~0.2 mL)) were intraperitoneally injected at a dose of 20 mg/kg in the mice. The control group were intraperitoneally injected with 200 μL of saline for each mouse. The mice were subsequently irradiated by 5 Gy gamma rays from $^{137}$Cs (photon energy 662 keV) with an activity of 3600 Ci p.i.. Forty-eight mice were assigned to the following six groups (eight mice per group): control, $Au_{10-12}(SG)_{10-12}$ only, radiation only, and $Au_{10-12}(SG)_{10-12}$ + radiation. Every group includes four male and four female mice in order to monitor the gender difference. The tumor sizes were measured every two or three days and calculated according to this equation: tumor volume = (tumor length) × (tumor width)$^2$ / 2.

**In vivo Toxicity.** Mice were purchased, maintained, and handled using protocols approved by the Institute of Radiation Medicine, CAMS. The mice treated with saline (control) and $Au_{10-12}(SG)_{10-12}$ nanomolecules only were weighed (Figure S2a) and assessed for behavioral changes. All mice were sacrificed at 20 days p.i., and their blood and organs were collected for hematology (Figure S3), biochemistry (Figure S4) and pathological investigation (Figure S5) from therapy mice. The blood was drawn for hematology analysis (potassium EDTA collection tube) and serum biochemistry analysis (lithium heparin collection tube) using a standard saphenous vein blood collection technique. During necropsy, the liver, kidney, spleen, heart, lung, brain, genitals, tumor, and thyroid were collected and weighed. Major organs including the liver, spleen, and kidney from these mice were then fixed in 4% neutral buffered formalin, processed into paraffin, and stained with hematoxylin and eosin (H&E). The pathology data (Figure S5) were collected with a digital light microscope.

## 2. Supporting Figures

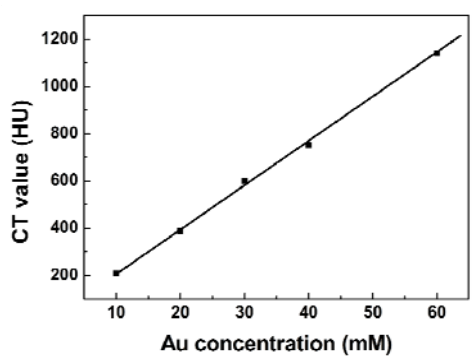

**Figure S1.** Experimental CT value of $Au_{10-12}(SG)_{10-12}$ nanomolecules as a function of Au concentration.

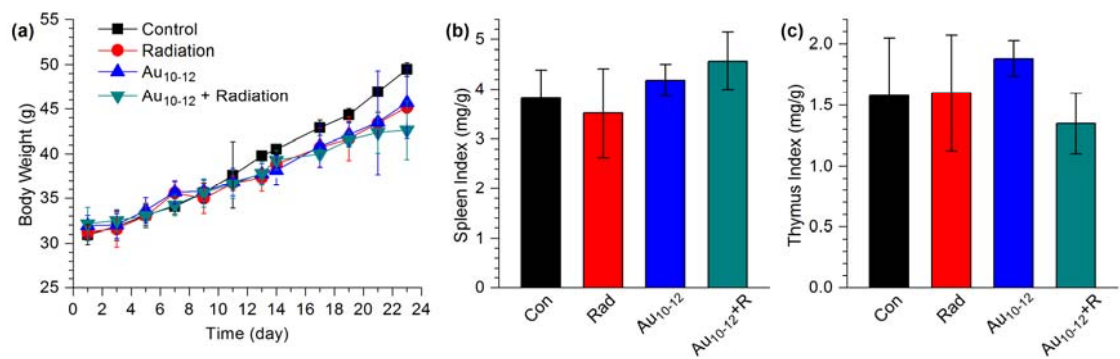

**Figure S2.** (a) Body weight, (b) spleen index and (c) thymus index of mice treated with saline only (control), radiation only, $Au_{10-12}(SG)_{10-12}$ nanomolecules (20 mg-Au/kg-body) only, and $Au_{10-12}(SG)_{10-12}$ nanomolecules (20 mg-Au/kg-body) + radiation.

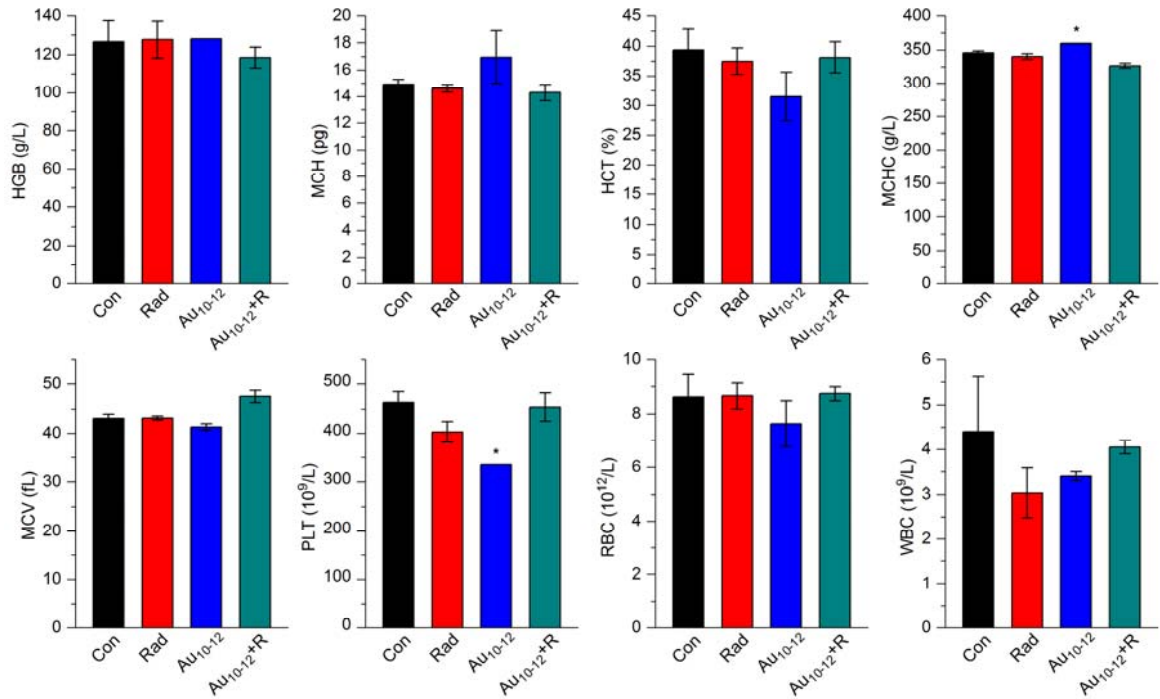

**Figure S3.** Hematology analysis of mice treated with $Au_{10-12}(SG)_{10-12}$ nanomolecules (20 mg-Au/kg-body) at 23 days p.i. The results show the mean and standard deviation of white blood cells (WBC), red blood cell (RBC), hematocrit (HCT), mean corpuscular volume (MCV), hemoglobin (HGB), platelets (PLT), mean corpuscular hemoglobin (MCH), and mean corpuscular hemoglobin concentration (MCHC). Data is analyzed by student's *t*-test, and the star represents significant difference from the control group ($p < 0.05$).

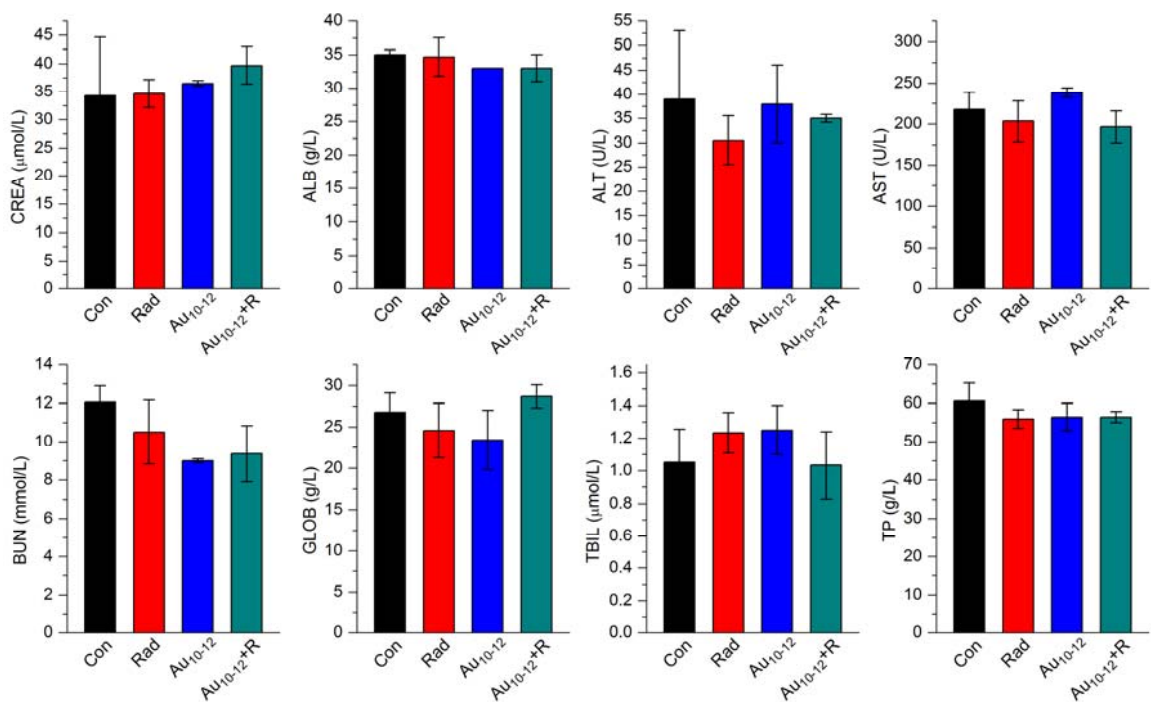

**Figure S4.** Blood biochemistry analysis of mice treated with $Au_{10–12}(SG)_{10–12}$ nanomolecules (20 mg-Au/kg-body) at 23 days p.i. The results show the mean and standard deviation of alanine aminotransferase (ALT), aspartate aminotransferase (AST), total protein (TP), ALB, blood urea nitrogen (BUN), creatinine (CREA), globulin (GOLB), and total bilirubin (TB).

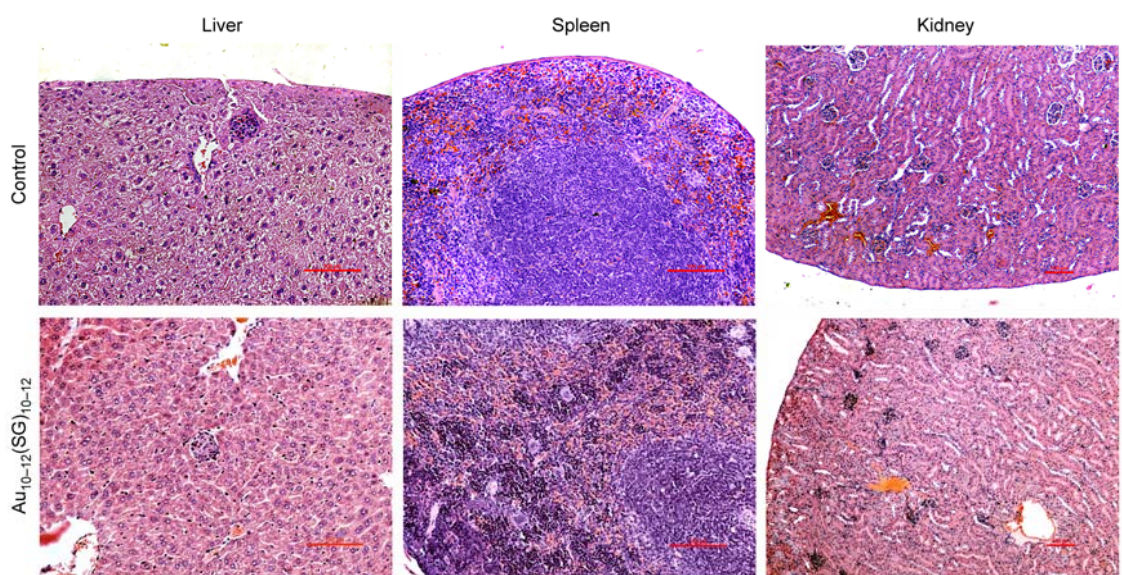

**Figure S5.** Pathological data from the liver, spleen, and kidney of mice treated with $Au_{10–12}(SG)_{10–12}$ nanomolecules (20 mg-Au/kg-body) at 23 days p.i.